\documentstyle[11pt]{article}
\begin{document}

\begin{flushright}
Liverpool Preprint: LTH 371\\
Edinburgh Preprint: 96/5\\
 hep-lat/9605025\\
 21 May 1996\\
 \end{flushright}
  
\vspace{5mm}
\begin{center}
{\LARGE\bf Orbitally excited and hybrid mesons from the lattice   }\\[10mm] 
{\large\it UKQCD Collaboration}\\[3mm]
 
{\bf  P. Lacock, C. Michael}\\

{Theoretical Physics Division, Department of Mathematical Sciences, 
University of Liverpool, Liverpool, L69 3BX, U.K.}\\[2mm]

{\bf P. Boyle, P. Rowland}\\

{Department of Physics \& Astronomy, University of Edinburgh, 
Edinburgh EH9 3JZ, Scotland}\\

\end{center}

\begin{abstract}

We discuss in general the construction of gauge-invariant non-local 
meson operators on the lattice.   We use  such operators to study  the
$P$- and $D$-wave mesons as well as hybrid mesons in quenched QCD, with
quark masses near the  strange quark mass.  The resulting spectra are
compared with experiment  for the orbital excitations. For the states
produced by gluonic  excitations (hybrid mesons) we find evidence of
mixing for non-exotic  quantum numbers. We give predictions for masses
of the spin-exotic hybrid  mesons with $J^{PC}=1^{-+},\ 0^{+-}$, and
$2^{+-}$.

\end{abstract}

\section{Introduction}

A quantitative determination of the hadronic spectrum of QCD has been
one of the major goals of physics for the past 20 years or so. A way to
determine the spectrum from first principles was first made possible in
lattice gauge theory, and  much  effort has been invested to obtain the
spectrum using lattice techniques. 

Up until now, most effort has gone into the determination of the masses
of states that are accessible through the use of spatially local
operators, i.e. states with quantum numbers that can be constructed
using  quark and antiquark propagators from a common point.  This is
mainly due to  practical considerations (e.g. limited computer-time
resources).  However,  extending  these studies to include states with
excited orbital momentum $L$ is of  importance in order to gain a more
complete insight and better understanding of the QCD spectrum.

From a theoretical point of view,  the  $L$-excited quark model gives a
very good qualitative description  of the observed meson spectrum.  The
low lying  mesons are the $S$-wave  $\pi$ and $\rho$  multiplets,  next
come the $P$-wave states: b$_1$ (B), a$_0$ ($\delta$),  a$_1$ (A$_1$)
and f$_2$ (f$_0$) multiplets, where we give the historic name  of the
isovector member of the multiplet in brackets.  Some of the $D$-wave
states are also recognisable: $\pi_2$ and $\rho_3$ (g).  This
classification incorporates most of the  prominent low-lying mesons. As
the orbital angular momentum $L$  is increased, linear Regge
trajectories are found with $m^2$  approximately linear in $L$.  States
which lie outside the naive quark model, glueballs and hybrid mesons,
are  not clearly seen  in experimental analyses of the spectrum.  It  is
important to understand these salient features of the meson spectrum
directly from  QCD. It is also  necessary to guide experimental searches
for hybrid mesons by establishing  the non-perturbative QCD predictions
for them.

Lattice gauge theory  provides a way to achieve this. Indeed the static
quark approximation  has provided information on both the
$L$-excitations and hybrid levels of heavy quark systems such as $c
\bar{c}$ and $b \bar{b}$ mesons. This programme of study has been
conducted in the  context of the static quark potential.  A
determination of spin-spin  and spin-orbit potentials has enabled the
fine and hyperfine structure  to be understood quantitatively. One of
the successes has been the lattice  identification~\cite{spin} of the
nature of the the long-range spin orbit  potential  which is responsible
for the correct description of the splitting of the $P$- and $D$-wave
mesons. Another important input from lattice QCD has been the 
determination~\cite{hybrid} of the ordering and energy of the hybrid
levels. This approach uses the Schr\"odinger equation  in the adiabatic
approximation and the prediction is that hybrid mesonic states will lie
close to the $D\bar{D}$ and $B\bar{B}$ thresholds. It is important for
experimental searches to  decide more accurately whether these  hybrids 
are above or below these thresholds. This requires a determination for
hybrid mesons of the effects of mixing with other levels and the effects
of   spin-orbit contributions. These are expected to be  quite large and
have not been evaluated by lattice methods.  As well as the static quark
 lattice work described, the effective lagrangian of NRQCD would also be
able  to explore  the hybrid spectrum for $b$-quark systems.

Here we wish to pursue these  studies of spectroscopy for lighter
quarks.  This is of great experimental interest since the heavy  quark
results described above are only of qualitative relevance to lighter
quarks. We shall proceed by using explicit quark propagators to 
construct mesonic states. Thus we shall directly obtain the masses  of
the mesons with the quantum numbers constructed. We show how   to study
all $J^{PC}$ states which allows an investigation of hybrid mesons in
this way.  

We make a first step  by  using the quenched approximation to explore
these states from first  principles.  The quark mass used here is
comparable to the strange quark mass.  We are able to explore the
$L$-excited spectrum for $P$- and $D$-waves.  We also explore hybrid
mesons and look explicitly at states with $J^{PC}$- values not allowed
in the quark model. These  {\it exotic} mesons are an  important signal
experimentally of gluonic excitations in the meson spectrum.

The $P$- and $D$-wave excited mesons have been explored previously using
either local operators (this only gives access to the $P$-wave $1^{+-}$,
$0^{++}$ and $1^{++}$ mesons)~\cite{ape,J,wingate} or by using explicit 
non-local operators~\cite{degh}. The work by DeGrand and
Hecht~\cite{degh} used  Coulomb gauge fixing and explicit $P$- and
$D$-wave non-local operators which were motivated by the spherical
harmonics. This  pioneering study~\cite{degh,Hechtetal} has been
extended to include a classification of the $P$-wave states under the
lattice symmetries~\cite{wingate}. Compared to previous work,  we are
able to get good signals  at lower quark mass, we have reduced finite
size effects and we are able to study a very wide range of states using
the full  classification under the symmetries of the lattice. The
additional advantage of our approach is that we can study this wide
range of quantum numbers with only two quark propagator inversions per
configuration.

The motivation for the construction of our mesonic operators is to 
follow similar ideas to those found to be successful~\cite{hybrid} for
the analysis  of the ground state potential and excited potential
between static quarks. Even though we work with much lighter quarks
(near the  strange quark mass) we expect similar methods to be
promising. Thus  we use non-local meson operators which are gauge 
invariant with a path of links joining the quark and antiquark. A
spatial fuzzing algorithm was  found to be very efficient for improving
the overlap with the ground state for  static quark studies~\cite{pot}
and, moreover,  using fuzzed links  was found to be
successful~\cite{fuzz} for the construction  of $S$-wave meson states.
Hence we choose to use fuzzed links  to construct the path here too.

The essence of our method is that at the source the quark is fixed at
$x_1=(0,0,0,0)$ while the antiquark is fixed at $x_2=(0,0,r,0)$. This
implies that only two quark propagator inversions are needed. One of
them is  the usual local propagator from  the origin and the other is
just the same local propagator but from a  local source shifted by $r$
in the 3-direction.  The choice of path to  join these sites is then
quite flexible - no additional quark matrix  inversions are needed for
different choices. We find that a straight  path is appropriate for a
study of $P$- and $D$-wave excitations, while a U-shaped  path gives
access to a study of hybrid mesons. We are then able  to construct and
measure correlations of these sources with operators at the sink that 
allow a wide range of quantum numbers to be studied. The  construction
and classification of these mesonic operators on a lattice is discussed
in detail in  the Appendix. In particular we show how, in principle,
mesons  with any $J^{PC}$ value can be studied using lattice methods.

In the next section we present our results for the $P$- and $D$-wave
excited  mesons. We then discuss the hybrid meson spectrum and
wavefunctions. In this work we have limited statistics so we concentrate
on presenting  the method and an analysis of the construction of the
most efficient  operators to use in such a study. We also give a first
estimate  of the mass spectrum using our methods.

\section{Orbitally excited mesons}

Our numerical results for quenched $SU(3)$ gauge theory are obtained
using quark propagators with local source at the origin  and a further
set of propagators in  the same configurations with local source 
displaced by $r$ units from the origin in the 3-direction (the choice of
spatial axes is arbitrary: one could just as well have chosen the 1- or
2- directions). 

Both sets of propagators were calculated on a $16^3 \times 48$ lattice
at $\beta=6.0$ using a tadpole improved clover action~\cite{ukqcd}. In
this exploratory study we  use one  hopping parameter value of 
$K=0.137$, which corresponds roughly to the strange quark mass. Thus  our
study directly bears on the $s \bar{s} $ spectrum.  In this first study,
we use 70 configurations and we set the source separation $r=6$. Our
results will enable us to decide if this was a good choice of $r$. Since
the static quark studies show that a  large fuzzing level is optimum, we
choose as our  fuzzing level to use 18  iterations with a coefficient of
the straight links $c=2.5$. We did  explore using a smaller fuzzing
level (5 iterations with $c=2$) and found  that our results were
essentially unchanged.

For the study of $S$-, $P$- and $D$-wave states, we use straight paths
to create the mesonic operators. For straight paths along the lattice
axes, this  allows a study of the states discussed in the Appendix. 
Here we choose to use the large component Dirac couplings
to the quarks (i.e. $(1+\gamma_4)/2$ projection).  We follow the
convention  of using the  historical name of the isovector meson to
label the state we are discussing - although it is the strange
iso-scalar that we are  effectively  creating.  To explore the  optimum
choice of operator, we use straight paths of length $R$ at  the sink and
sum over all spatial sites to have zero momentum. The dependence of the
mesonic correlation on $R$ gives  direct information on the Bethe
Salpeter wavefunction of the meson.  This has already been studied by
essentially our present methods for the $S$-wave mesons ($\pi$ and
$\rho$)~\cite{fuzz,gupta}. 

 To  study the mesons most accurately, we use as many operators as
possible in order to  make the most constrained fit to the measured
correlations versus $t$. Thus we can use results for several values of 
$R$ together. Moreover, we can include correlations from additional
operators for those $P$-wave mesons which can be created by the small
Dirac components of the quark bilinear  (namely $\delta$, A$_1$ and B)
together with an isotropic spatial operator. Since we have access to
local-local  and fuzzed-local correlations from 375
configurations~\cite{ukqcd}, we are able to use these data to improve
the mass determination substantially  for those three mesons.

Because the correlations decrease rapidly with increasing time
separation, we  choose to fit to as low a $t$-value as possible with a
two state fit (note that we only measured the shifted-source
correlations up to $|t|=10$). Fitting simultaneously as many different 
types of correlation as possible,  an accurate determination of the
ground state mass is obtained since the approach to a plateau is well
modelled.   For the  $\delta$, A$_1$ and B, we used $P$-wave operators
with  sizes $r=6$ at the source and $R=6$ and 2 at the sink, and
$S$-wave operators using the small Dirac components with sizes $r=0$ and
6 at the source and $R=0$ at the sink. Thus we fit data from 7 different
correlations  of combinations of source and sink $P_6 P_6$, $S_0 P_6$, 
$P_6 P_2$, $S_0 P_2$,  $P_6 S_0$, $S_0 S_0$ and $S_6 S_0$. The effective
masses for these correlations are illustrated for the B meson in Fig.~1.
From the constraints of factorisation, we deduce that at least  two
excited states (ie a three state fit) would be  needed to describe all
the  data down to $t=3$  because the $P_6 P_6$ and $S_0 S_0$ effective
masses show excited state contributions  which are not present for $S_0
P_6$. Our  fitting strategy is to determine the ground state mass by
fitting  the largest sample of data simultaneously,  consistent with an
acceptable estimate for $\chi^2$.  This $\chi^2$ estimate comes from
modelling the correlations among  the data to be fitted by retaining the
8 largest eigenvalues and averaging the remainder~\cite{mm}.  The
statistical errors on these fits are taken  from a full bootstrap and  a
systematic error is estimated from varying the fit region in $t$, the
data set fitted, and the type of fit. We illustrate in Fig.~1 a
non-factorising fit with $t \ge 3$ for the B meson.  Similar results for
 the ground state mass are obtained from a factorising fit  with $t \ge
4$. The ground state mass values are shown  in Table~1.

\begin{figure}[p]
\vspace{14cm} 
\includegraphics{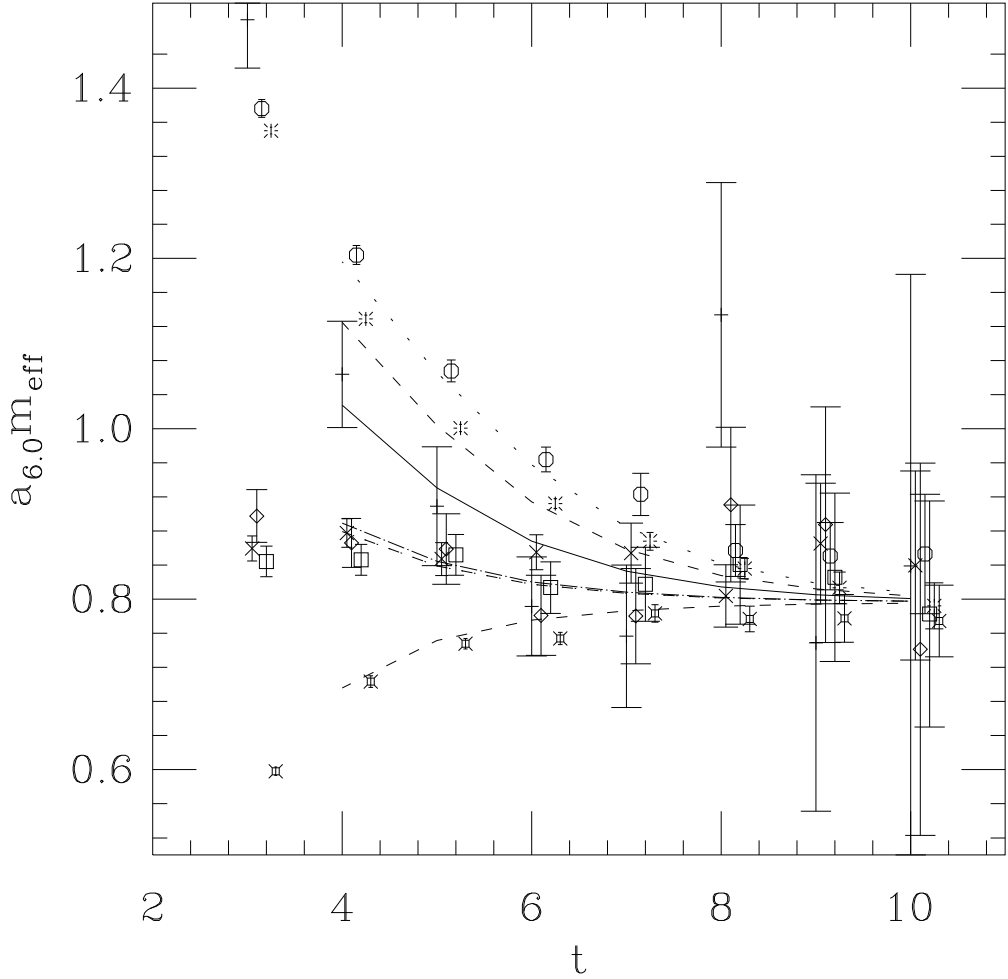}
 \caption{ The two state fit to the effective mass for the $1^{+-}$
meson  versus time $t$.  We used $P$-wave operators with   sizes $R=6$ and
2, and $S$-wave operators using the small Dirac components with sizes
$R=0$ and 6. Effective mass results shown are from correlations  of
combinations of source and sink $P_6 P_6$  (+), $S_0 P_6$ ($\times$),
$P_6 P_2$ ($\diamond$), $S_0 P_2$ (octagon),  $P_6 S_0$ ($\Box$), $S_0
S_0$ (*) and $S_6 S_0$ (fancy square).  
   }
\end{figure}

For the f$_2$ meson, we have access to it via two different  cubic group
operators. Fits to these operators separately show that they are
consistent  with having equal correlations. Our results are presented
from fits assuming this. In this case  only $P$-wave operators  are
available,  so we use a simultaneous fit with  $R=6$, 4 and 2. The
ground state mass is given  in Table~1.

\begin{figure}[p]
\vspace{14cm} 
\includegraphics{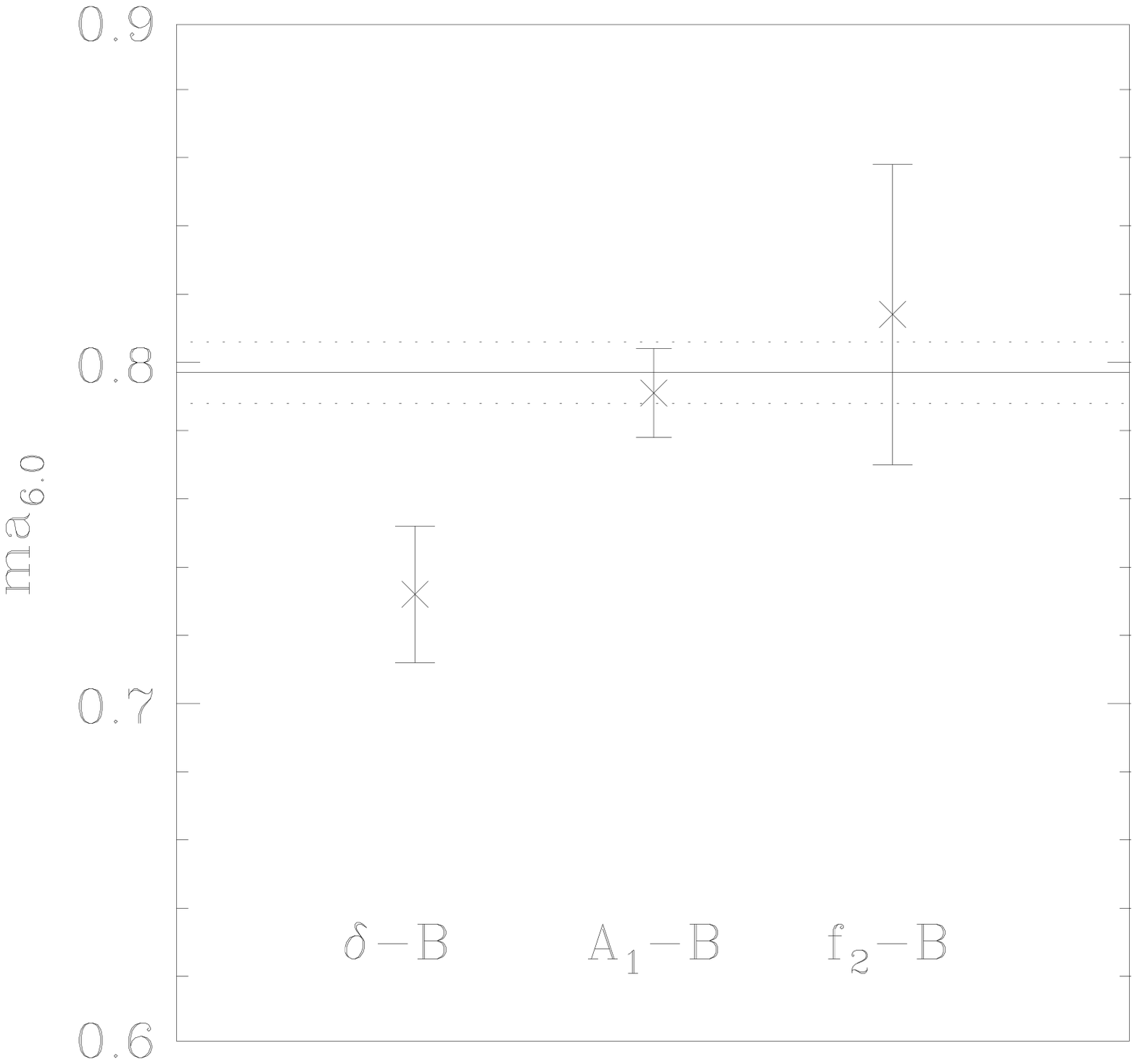}
 \caption{ The differences of the mass (in lattice units) for the 
$P$-wave triplet states relative to the singlet.
   }
\end{figure}

The results, shown in Fig.~2 and listed in Table~1, suggest that the
f$_2$, A$_1$ and B members of the $P$-wave multiplet are degenerate
within errors. The $J=0$ state, $\delta$, is consistent with being
lighter than the other $P$-wave states by $3 \sigma$  where,  in order
to investigate the fine structure splitting in the  $P$-wave meson
multiplet, we perform correlated $\chi^2$ fits to the data using the
same bootstrap samples and $t$ interval (3 -10) - the results are shown
in Fig.~2.

The $D$-wave mesons with $J^{PC}=2^{-+}$ and $2^{--}$ can be   studied 
using the straight paths along the lattice axes. We make similar fits as
for the f$_2$ meson (using two exponentials to fit simultaneously $R=6$,
4 and 2).  A typical fit (to the $2^{-+}$ meson) is shown in Fig.~3. 
Our results for the masses using straight paths are shown in Table~1. 
However, the straight-path  operator does not allow the $D$-wave mesons
with $J^{PC}=2^{--}$ and $3^{--}$ to be separated.  As discussed  in the
Appendix, L-shaped paths can be used to achieve this. We explored this
using local sources at $(0,0,0,0)$ and $(6,0,6,0)$ and were able  to
determine the masses separately. The results, also shown in  Table~1,
from 20 configurations are consistent with these two states being
degenerate  within our errors.

\begin{figure}[p]
\vspace{14cm} 
\includegraphics{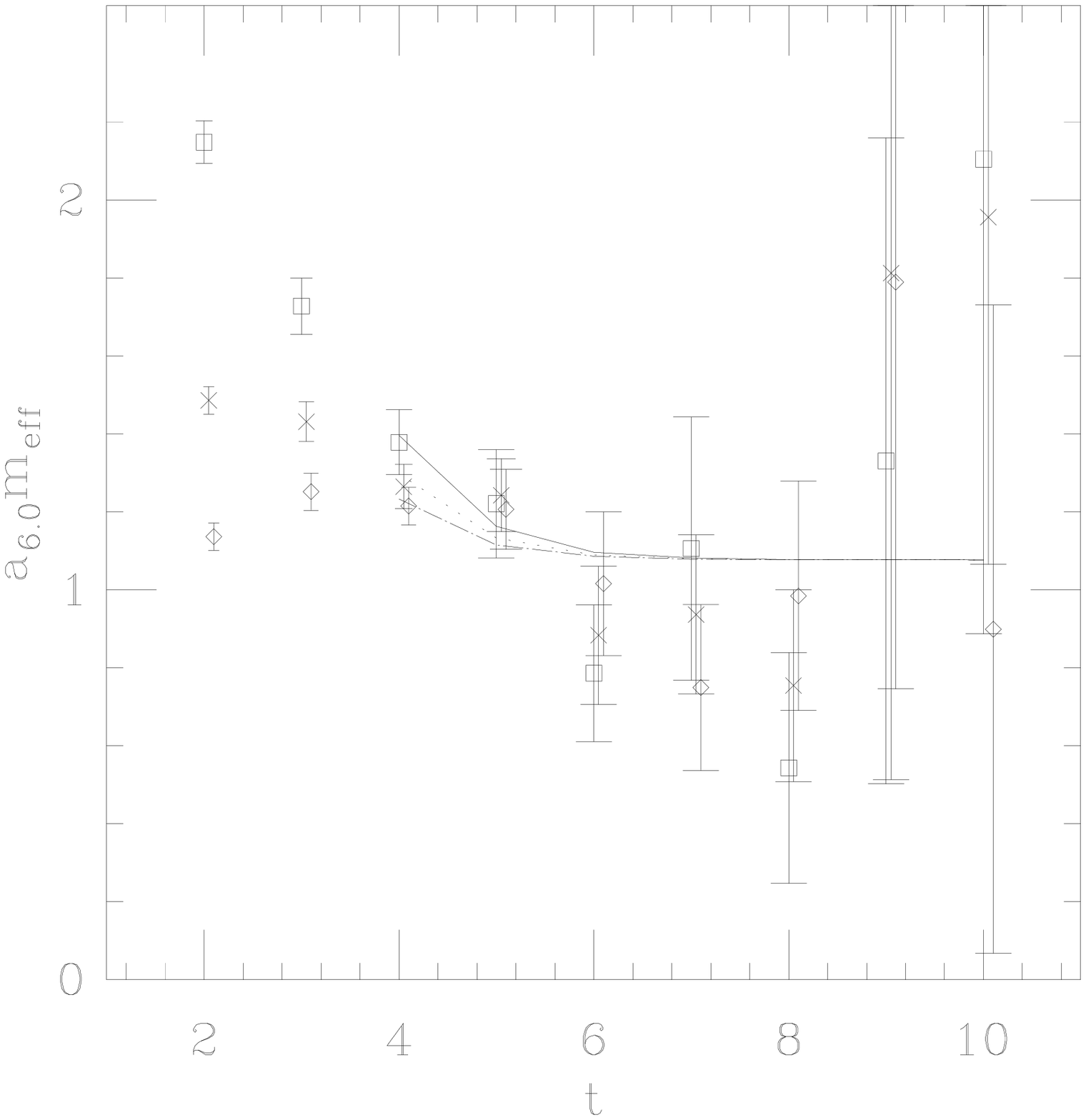}
 \caption{ The two state fit to the effective mass for the $2^{-+}$
meson  versus  time $t$. Correlations for sink sizes $R=2$ ($\diamond$),
4 ($\times$)  and 6 ($\Box$) are shown, where the source has $r=6$.
   }
\end{figure}

For the vector meson, there are both $S$-wave and $D$-wave couplings. We
made a simultaneous fit to the  $S$-wave and $D$-wave operators (10
types of correlation in all of which the $S$-wave local-local  and
fuzzed-local correlations come from 375 configurations) to determine the
first excited state. The ground state was found to couple weakly to the
$D$-wave operator  whereas the excited state  was  found to couple
strongly to the $D$-wave operator but weakly to the $S$-wave  operator
for $R=6$. The latter fact is a confirmation of our previous~\cite{fuzz}
observation that the excited state had a node at $R=6$. The systematic
errors on the excited state determination  are relatively large because
the ground state dominates at large $t$. Our  result with statistical
errors only is given in Table~1.

Our results for $P$- and $D$-wave mesons are considerably more accurate 
than those obtained by ref~\cite{degh} for light quarks. Indeed,  for
the first time, we are able to see a signal for the splitting of the
$P$-wave  multiplet.

\begin{table}
\begin{tabular}{llccllc}
& meson&  $J^{PC} $ & t-range  & mass&m/m($\phi$)&expt($s\bar{s}$) \\
& multiplet &      &         & $ma$ &    &GeV  \\
$L=0,4$&& & &  \\
&$ \pi$&$  0^{-+},\ 4^{-+}$ &5,23 &0.413(1)&0.76(1)& 0.68  \\
&$\rho$&$   1^{--},\ 3^{--}$ &5,23 &0.540(2)& 1.00&1.02 \\
$L=1,3$& & & & &&\\
&b$_1$ (B)&$  1^{+-},\ 3^{+-} $ &3,10 &0.797(8)(10)&1.48(2)& 1.42 \\
&a$_0$ ($\delta$)&$ 0^{++},\ 4^{++}$ &3,10 & 0.737(16)(10)&1.36(3)& 1.21 \\
&a$_1$ (A$_1$)&$  1^{++},\ 3^{++}  $ &3,10 &0.789(8)(10)&1.46(2)& 1.43 \\
&f$_2$&$  2^{++},\ 3^{++}$  &3,10 &0.817(37)(30)&1.51(7)& 1.52 \\
&f$_2$&$  2^{++},\ 4^{++} $  &3,10 &  \dots &\dots &\dots \\
$L=2,4$ & &  & &&\\
&$ \pi_2$&$    2^{-+},\ 4^{-+} $  &3,10 &1.07(15)(15)&1.98(28) &1.81 \\
&$ \rho$&$    1^{--},\ 3^{--} $ &3,10 & 0.543(3)&1.01(1) &1.02\\
 &$\rho'$ & & &1.25(13) &2.31(24)&1.68\\
&$\rho_2$,\ $ \rho_3$&$   2^{--},\ 3^{--} $ &3,10&1.19(6)&2.20(11) & \\
$L=2,3$ & &  && &&\\
&  $\rho_3$\ (g)&$   3^{--},\ 5^{--} $ &2,8 &0.86(19) &1.59(35) &1.85\\
&$\rho_2$ &$     2^{--},\ 4^{--} $ &2,8 &0.79(12) &1.46(22) &\\
\end{tabular}
 \caption{The masses of the mesons with $S$-, $P$- and $D$-wave
excitations (with $s\bar{s}$ quarks but labelled by the name of the I=1
particle in the multiplet).  The $J^{PC}$ values are the lowest two
values allowed by the lattice cubic symmetry. The angular momentum $L$
of the operator is shown similarly.  For those $P$- and $D$-wave mesons
which are comparatively well determined in  mass, the second error shown
is an estimate of the systematic error from different fits.  The last
two rows are from `diagonal straight  paths' as described in the text
(with lower statistics). More details of the construction of the
operators used  are given in the Appendix. The experimental masses of
the $s \bar{s}$ states are given,   as an educated guess in some
cases (see Sect. 4).}
 \end{table}

The extraction of the ground state as outlined above also  allows the
wavefunction to be obtained as $R$ is varied. Results  are given for the
$S$-, $P$- and $D$-wave spin-singlet states  in Fig.~4.  These show
clearly the expected functional  behaviour with $R$  of  $R^L$ for $S$-,
$P$- and $D$-wave mesons. Our use of a gauge-invariant path  description
allows us to probe this wave function out to beyond  half the spatial
lattice size - unlike the case where a gauge-fixed method is
used~\cite{Hechtetal} - results for $R>6$ were  obtained with less
statistics (20 configurations) and show a plausible  behaviour with no
disruption from finite spatial size effects.

\begin{figure}[p]
\vspace{16cm} 
\includegraphics{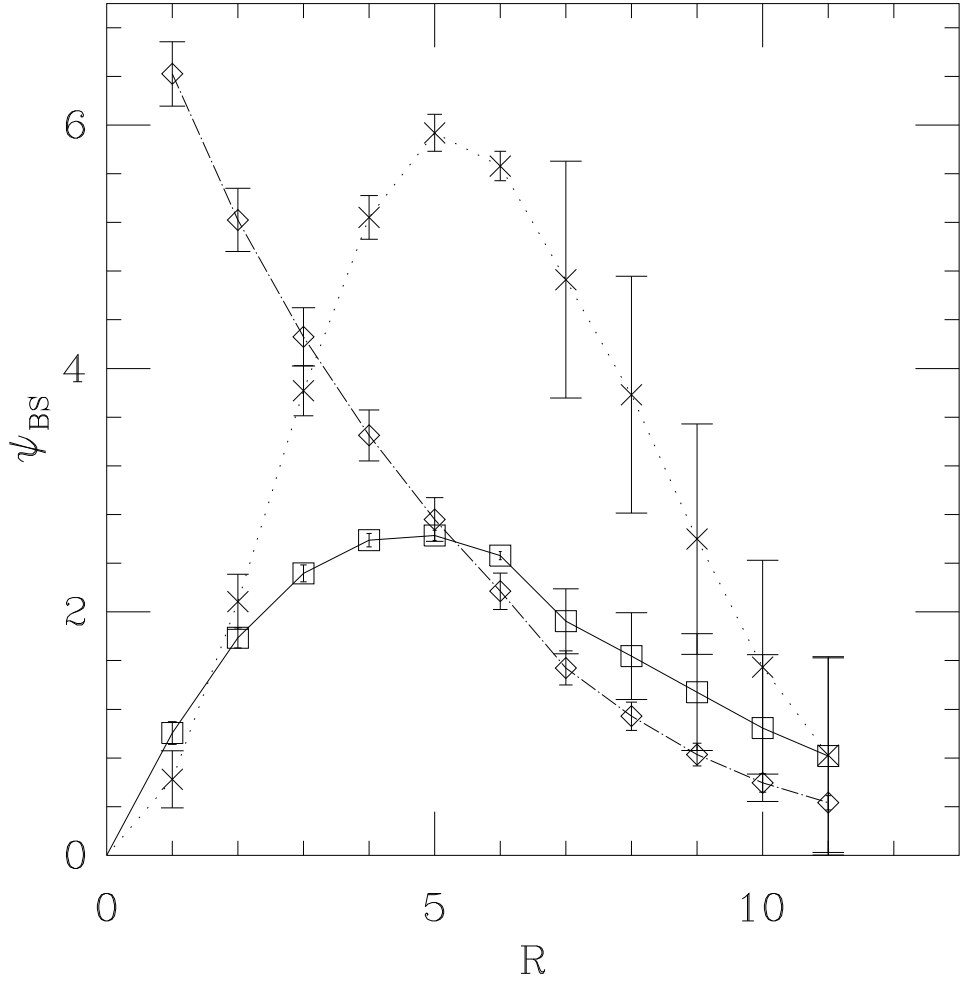}
 \caption{ The Bethe Saltpeter wavefunctions obtained from our fuzzed
straight paths of length $R$ for $S$- ($\diamond$),  $P$- ($\Box$) and
$D$-wave ($\times$) spin singlet mesons. The results for $R>6$ have
smaller statistics. Here $R$ is in lattice units ($a \approx 0.5
$GeV$^{-1}$). The normalisations of the wavefunctions are arbitrary.
   }
\end{figure}

Wavefunctions for the $P$- and $D$-wave mesons have been obtained by
DeGrand and Hecht~\cite{degh}. The Bethe Salpeter wavefunction depends
on the method used to construct it.  Their method uses a Coulomb gauge
fixing and so is not  directly comparable to our determination.
Qualitative features are in  full agreement, however. The $D$-wave
$\rho$ operator has also  been studied~\cite{gupta} by a method similar
to ours and the result  is consistent.

Our choice of $r=6$ to use as a source separation is seen to be a
reasonable  compromise for the studies at
different orbital momentum $L$. The original
choice of $r=6$ was  motivated by the discovery that the radial
excitation of the $S$-wave mesons  has a node at this value~\cite{fuzz}.
The orbital  excitations have no such node as we see in Fig.~4. Since
we  measure correlations for different $R$ at the sink, we can 
construct optimum operators (in a variational sense that any excited
state is  minimised) at the sink.  

Our choice of source operator is a `white source'. It has
contributions  to nearly all quantum numbers. This is an advantage in
allowing a comprehensive  study of the meson spectrum without extra
quark matrix inversions. It is a  disadvantage, however, in that the
signal in any given quantum number  will be noisier because of the poor
overlap of the source operator  with that quantum number. This dilemma
was discussed by DeGrand and Hecht~\cite{degh}  who concluded that a
rather high symmetry was needed at the source. We  find that the fuzzed
construction allows our rather `white' source  to give a reasonable
signal to noise ratio. Of course it could be improved for  any
particular state by using  a quark propagator from a more symmetric
fixed  source  (e.g. a suitable sum/difference of  sources  at $(0,0,\pm
r)$, $(\pm r,0,0)$ and/or $(0,\pm r,0)$). The increase  in signal to
noise would be compensated by decreased flexibility since only  specific
quantum numbers would be created. A compromise is possible, namely
retaining flexibility by using  inversions from several sources
separately so that they can be combined with different coefficients as
needed. However, because propagators from sources that  are near to each
other are strongly correlated to each other, it may be  more efficient
to increase the number of configurations studied rather  than construct
such multiple sources.

\section{Hybrid mesons}

The lowest lying hybrid levels for static quarks~\cite{hybrid} come 
from an interquark potential in which the colour flux from quark to
antiquark is excited in a transverse  spatial plane as $\, \sqcap -
\sqcup$ (cf. Appendix for notations). Excitations of this kind are
clearly  non-trivial gluonic contributions and the mesonic states in 
such excited potentials include exotic $J^{PC}$ values.  As discussed in
more detail in the  Appendix, the same states can be accessed using
propagating quarks  by using a non-local mesonic operator in which the
quark and  antiquark are joined by such a difference of paths. This
construction  has the advantage that it does not require any more quark
propagator  inversions than those needed in the above study of $P$- and
$D$-wave  mesons. Indeed, since the U-shaped paths can be transverse in
either of  two spatial directions, we have more operators available.
Using eigenstates  of the lattice symmetry at the sink, we are able to
explore a wide range  of quantum numbers of interest.

We first address the issue of creating optimum operators (in the sense 
of best signal to noise for the ground state of a given quantum number).
Even with our fixed choice of $r=6$ in the 3-direction as source
separation,  we can vary the transverse extent $d$ of the U-shaped path
at the  source   and we can explore at the sink a range of values of $R$
and $D$  describing the U-shaped paths there.  We expect the signal to
increase  with the area ($R\times D$) of the U-shaped path but the noise
will  increase too. The static quark results~\cite{hybrid} suggest that
the  hybrid mesons have very extended wavefunctions in the $R$-direction
while the best signal comes from $D \approx 2$ for $R=6$. Another clue 
is that, for $S$-wave mesons made from propagating quarks, a smaller
sink extent gives a less noisy  signal than a smaller source size.

To investigate the optimum operators,  we used the exotic state with
$J^{PC}=1^{-+}$ as a probe and varied the source and sink operators to
maximise the  signal to noise ratio at $t=2$ and 3. There is an
advantage in using the  same operator at source and sink because the
effective mass is  strictly an upper bound in this case. The best signal
(i.e. lowest effective mass within  errors from  the ratio of $t=3$ to 2)
for the effective mass  for such a diagonal case came with
$(r,d)=(R,D)=(6,6)$ although the (6,3) case was very similar. For
off-diagonal correlations, the optimum signal to noise for the
correlation at $t=3$ was for $(r,d)=(6,6)$ and $(R,D)=(1,1)$. Taking
into account the above, we consider the combinations  $d$=6;   \  $(R,D)
= (6,6),\  (3,3)$ and  (1,1)  in the analysis below.

Since the construction of correlators involving the  hybrid meson
operators can be quite  computer-time intensive, we choose to include
only the  large components of the Dirac propagators, that is we
effectively  multiply the correlators with a factor of (1+ $\gamma_4$)/2
(cf. Appendix). Compared  to the results obtained in  the standard case
of summing over all four components of the  Dirac spinor we find that
the signal, and thus  both the correlation averages and errors, are
smaller, but the physical results (e.g. ratios like the effective
mass) are consistent in the two cases. However, even though there is no
gain in statistical accuracy, the resulting speed up of  the analysis
code is quite substantial. Similar behaviour is found for the  $P$- and
$D$-wave case.

For the hybrid operators we fitted simultaneously the three sink sizes 
discussed above and found acceptable fits for the $t$-range 2-10, using 
two exponentials.  To stabilise the fits we fixed the mass difference to
the first excited state at 1.5 in lattice units.  Statistical errors
were estimated  by jackknife. To establish the  systematic errors to the
ground state masses determined by the fits, we varied fits by allowing
the  minimum $t$-value to be 3 or by allowing the first excited state
mass to be free. In Figs.~5-7 we show the effective mass plot for the 
$J^{PC}=1^{-+},\ 0^{+-}$ and $2^{+-}$ exotic states with the fits. Note
that the mesonic correlation data was fitted directly (accounting for
the correlations among data points in the same way  as for the fits to
the orbital excitations) rather than the effective mass, which is
plotted. The effective mass plot  tends to mis-represent the errors when
the errors are large. Even though the error bars are indeed quite large,
the fits indicate that an acceptable plateau is reached for the
$t$-interval  considered.   The results for the ground state  masses
obtained from the fits  for the hybrid states are summarised in Table~2
and Fig.~8. The systematic errors on the ground  state masses are
consistent  with being smaller than the quoted statistical errors.

\begin{table}
\begin{tabular}{llccl}
& meson&  $J^{PC} $ & t-range  & mass\\
& multiplet &      &         & $ma$\\
$L=1^{+-}$&&&&   \\
&$\rho$&$   1^{--},\ 3^{--}$ &2,10 &0.53(9) \\
&$ \pi$&$  0^{-+},\ 4^{-+}$ &2,10 &0.40(2) \\
&$ \pi_2$&$    2^{-+},\ 4^{-+} $  &2,10 &1.56(76) \\
&$ \pi_2$&$    2^{-+},\ 3^{-+} $  &2,10 &1.40(53) \\
&$\hat{\rho}$&$  1^{-+},\ 3^{-+} $ &2,10 &0.99(28) \\
$L=1^{-+}$&&&&   \\
&A$_1$&$  1^{++},\ 3^{++}  $ &2,10 &0.99(56) \\
&\^a$_0$&$ 0^{+-},\ 4^{+-}$ &2,10 & 1.05(18) \\
&\^a$_2$&$  2^{+-},\ 4^{+-} $  &2,10 & 1.29(44) \\
&\^a$_2$,\ b$_3$&$  2^{+-},\ 3^{+-}$  &2,10 &1.17(26) \\
&b$_1$\ (B)&$  1^{+-},\ 3^{+-} $ &2,10 &0.84(31) \\
\end{tabular}
 \caption{The masses of the hybrid mesons created by operators with
spatial excitation  $L^{PC}=1^{+-}$ and $1^{-+}$.  The $J^{PC}$ values
are the lowest two values allowed by the lattice cubic symmetry.  More
details of the construction of the states are given in the Appendix. The
masses are given in lattice units and are appropriate to  the $s
\bar{s}$ states}
 \end{table}

The spin-exotic mesons cannot mix with $q \bar{q}$ mesons and so the 
ground state masses will be relevant to spectroscopy. The lattice 
values we find all have large errors and are compatible with being
equal.  The mass in physical units corresponds to 1.9(4) GeV. A further
increase  in precision will follow from the increase in statistics
planned. Even  our present statistical sample allows some estimate of
possible splittings  among these exotic states. The commonly held view
is that the $J^{PC}=1^{-+}$  state will be the lightest exotic and our
results are consistent with this.

\begin{figure}[p]
\vspace{14cm} 
\includegraphics{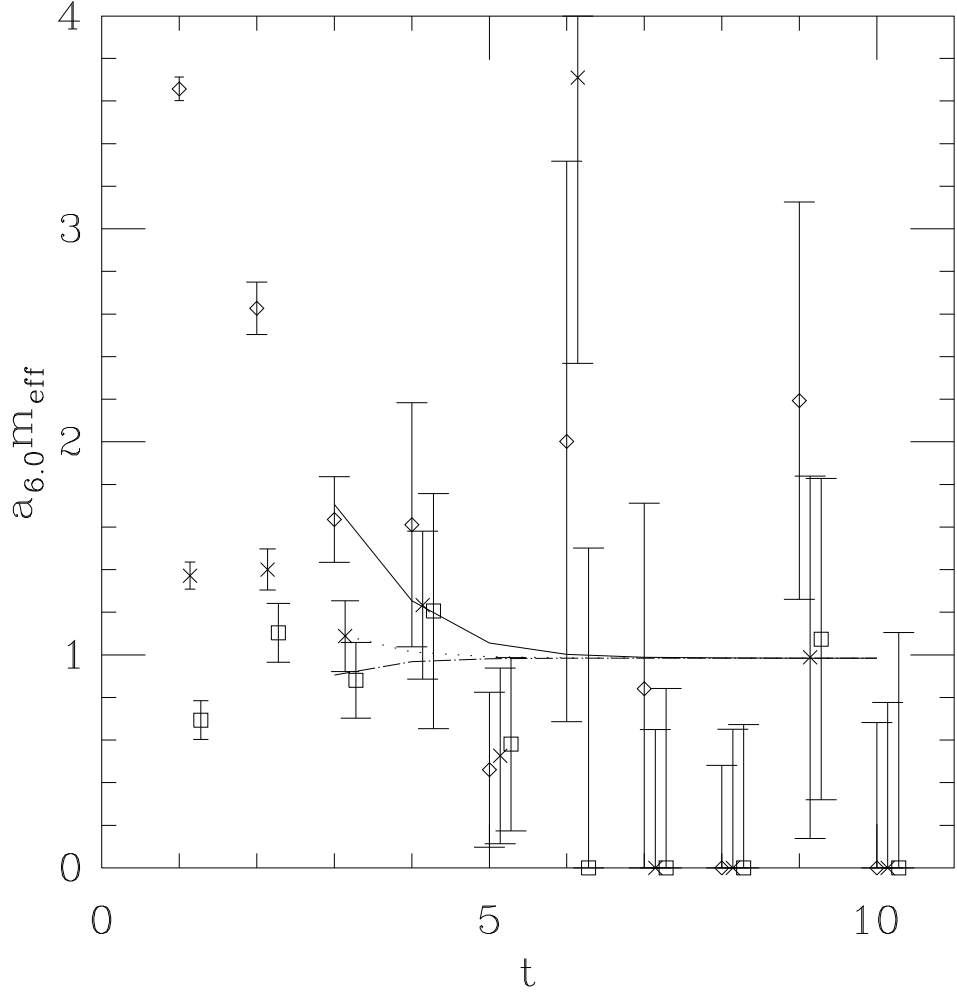}
 \caption{ The lattice effective mass for the $J^{PC}=1^{-+}$  hybrid
meson versus time separation $t$. The source used was a U-shaped path of
size  $6 \times 6$, while the sinks were combinations of U-shaped paths 
of size $6 \times 6$ ($\diamond$), $3 \times 3$ ($\times$) and $1 \times
1$ ($\Box$). The two state fit is shown.
   }
\end{figure} 

\begin{figure}[p]
\vspace{14cm} 
\includegraphics{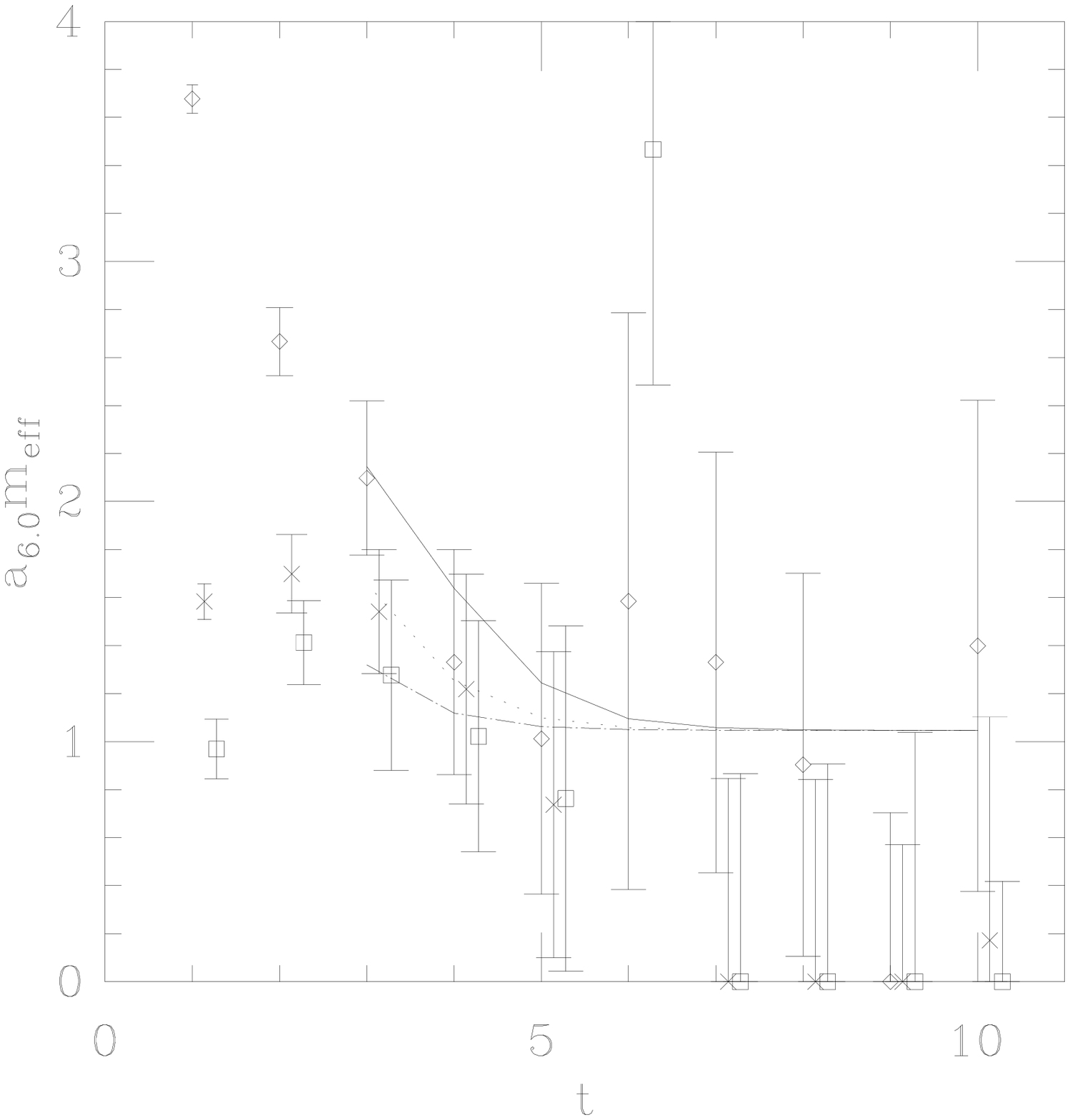}
 \caption{ The lattice effective mass for the $J^{PC}=0^{+-}$  hybrid
meson versus time separation $t$. Other details as Fig~5.
   }
\end{figure}

\begin{figure}[p]
\vspace{14cm} 
\includegraphics{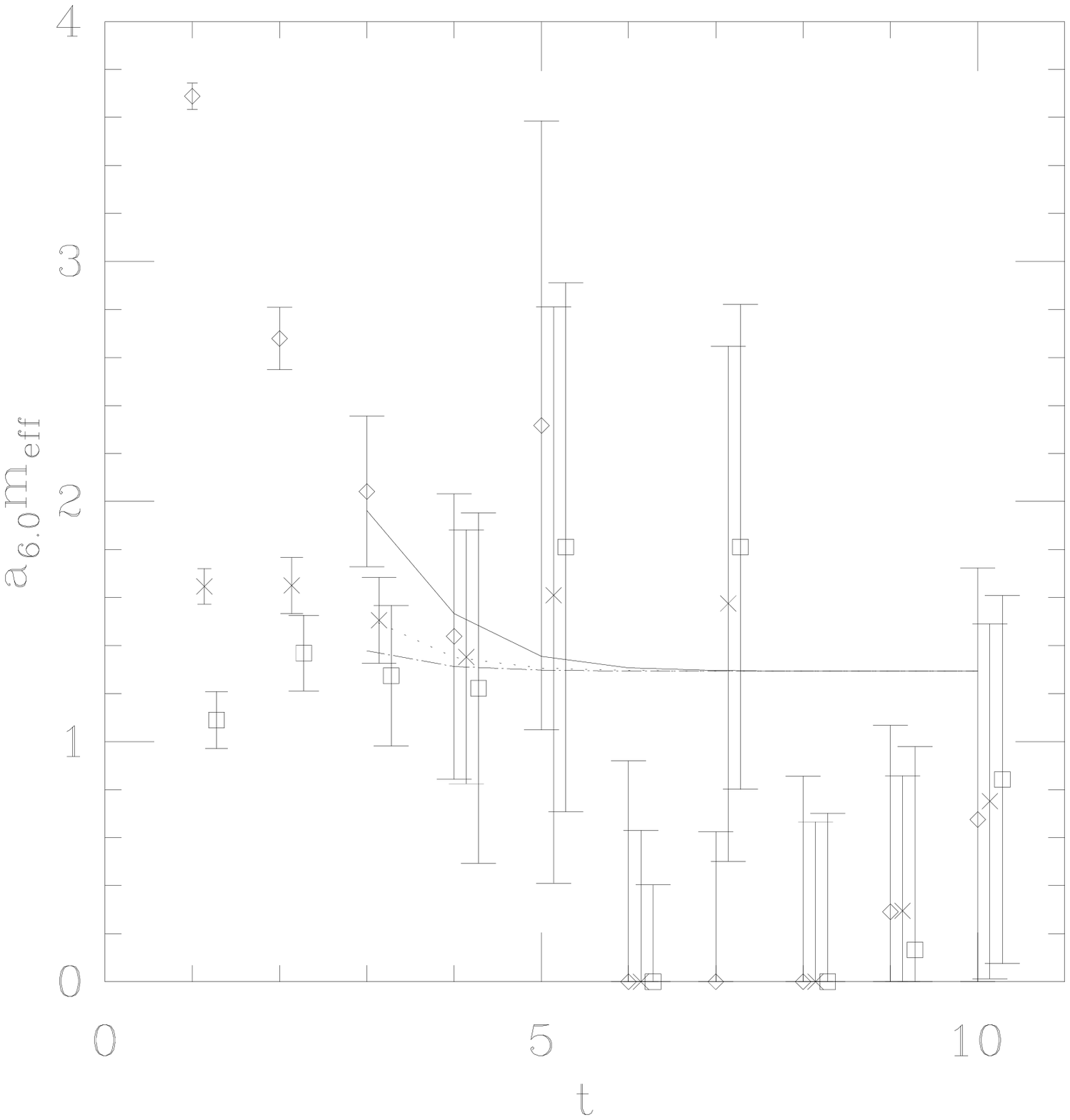}
 \caption{ The lattice effective mass for the $J^{PC}=2^{+-}$  hybrid
meson versus time separation $t$. Other details as Fig~5.
   }
\end{figure}

\begin{figure}[p]
\vspace{14cm} 
\includegraphics{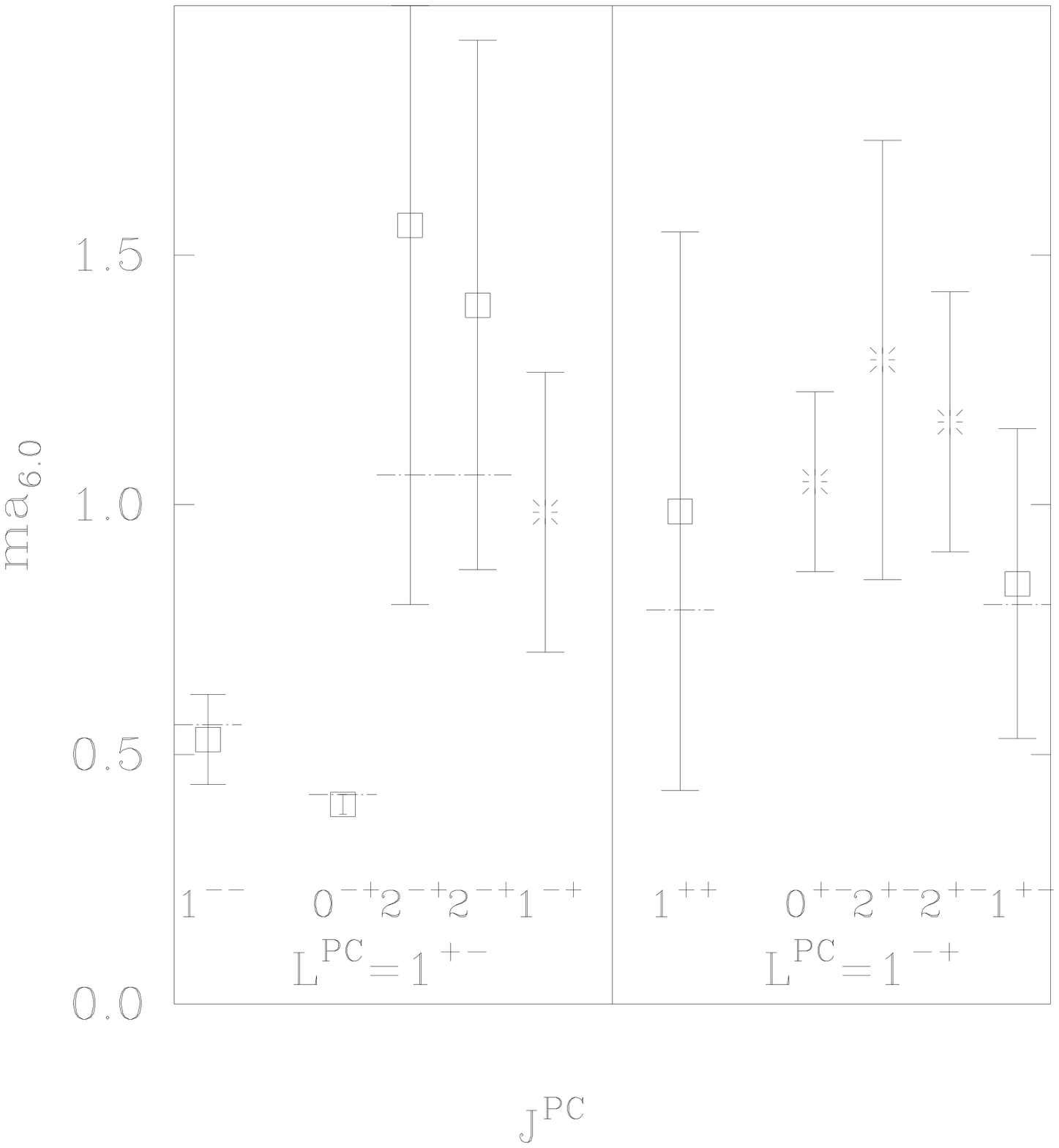}
 \caption{  Results for the ordering of the hybrid  meson
levels for strange quarks. The states with  burst
symbols are $J^{PC}$ exotic. The dashed lines represent $L$-excited
quark model states as determined on the lattice. The strong mixing of
the states  created by our hybrid operators  with these is apparent for
the pseuodoscalar and vector meson  cases.
   }
\end{figure}

In principle, states with the same $J^{PC}$ and different internal
spatial structure  will mix in quantum field theory.   Here we are  able
to explore these mixing possibilities in full by looking directly at the
meson spectrum  with propagating quarks. Fig.~8 clearly shows that
mixing of the hybrid states with non-hybrid ones for vector and
pseudoscalar  mesons is quite significant. Indeed our hybrid operators
for these states have a  substantial overlap with the $q \bar{q}$ ground
state mesons. The quantitative strength of  this  mixing is related to
the lattice operator construction we use. Relating this to Minkowski
wavefunction  mixing is not straightforward. The subjective impression,
however,  is that the mixing is larger that expected. Moreover, in the
case of the ground state vector meson (i.e. $\phi$), we can compare the
surprisingly small mixing of the $D$-wave operator found above with  the
large mixing of the hybrid operator found here.  This suggests that
gluonic excitations mix more readily  than orbital excitations.

The $0^{+-}$ exotic meson can be studied in principle using an 
isotropic spatial operator with a  $\gamma_4$  coupling between quark
spins.  With our larger statistics of 375 configurations for local-local
and fuzzed-local meson correlations, we explored this using $\gamma_4$
at source and sink but found no  signal.  An investigation of
correlations with a hybrid operator at the source (from 20
configurations  with L-shaped paths)  and a $\gamma_4$ sink also showed
no clear signal with this statistical sample.  These negative results 
could be explained if the spin-exotic state was indeed a hybrid with
excited gluonic fields and so would have a small  overlap with the
isotropic operator considered here which has the gluonic fields in their
ground state.

We have also investigated the possibility of using a  gluon flux loop to
construct the meson operator  with an excited gluonic component, but
with the quark and antiquark at the same site. This would allow
correlations  involving the operator to be  explored using local
propagators from one site only, which  avoids the need for extra
shifted-source inversions when  studying hybrid states.  The gluon flux
loop should have similar features to that used in the heavy quark case
for a static adjoint source ~\cite{gluelump}:  for example  an
anti-symmetrised  clover shape (i.e. clover paths minus their hermitean
conjugate) which has the quantum numbers corresponding to a magnetic
gluon $L^{PC}=1^{+-}$. The electric gluonic excitation can  be created
easily too - see ref~\cite{gluelump}.  Thus, using different gamma
matrix combinations, all hybrid states can, at least  in principle, be
produced at source and sink. We studied several cases and were able to
extract a signal from such closed-loop gluonic excitations for  a vector
meson treated as a  hybrid (using a closed magnetic gluon loop with
singlet quark spin combination). However, since we have already found
large mixing between hybrid operators and the  vector meson ground
state, this case does  not confirm that, in general, such a closed loop
gluonic excitation  is a useful way to create hybrid states. This lack
of success is perhaps a consequence of  the hybrid wavefunction being
small for quark-antiquark separation of zero - as is known to be the
case for heavy quarks~\cite{hybrid}.

\section{Discussion}

The mesonic spectrum in the quenched approximation has only been studied
in detail on the lattice  for some of the lowest states. Here we extend
this study to considerably more states.  The quenched approximation does
not include effects from mixing of the $q\bar{q}$ system with 
$q\bar{q}$ of  different flavour quarks, with $q\bar{q}q\bar{q}$ or with
purely gluonic components. Thus a `magic mixing' is expected  in the 
quenched approximation with the  $s\bar{s}$, $s\bar{q}$ and $q\bar{q}$
members equally split in  $m^2$, where $q$ means $u$ or $d$.  For the
$J^{PC}=1^{--},\ 2^{++}$ and $3^{--}$  mesons, such a mass spectrum is
close to that observed and we compare our results with  the I=0 state
assigned as $s\bar{s}$ ($\phi$,\ f$_2'$ and $\phi_3$ respectively). 

Within the quenched approximation, one can study disconnected  quark
loops to estimate the strength of the mixing efects. For the 
pseudoscalar mesons, this approach~\cite{kt} gives a reasonable
explanation of the  $\eta$ mass.  Because these disconnected loop 
contributions affect I=0 only,  we estimate  the mass of the
pseudoscalar $s\bar{s}$ state as 0.68 GeV, assuming magic mixing, from
the experimental masses of I=1  and I=1/2 states.
 
Charge conjugation C is not a good quantum number when the quark and
antiquark  have different masses. This implies that, for  the  axial and
$2^{-}$ mesons, the strange partners with C=$\pm 1$ mix.   For the
$1^{++}$ meson we identify the $s\bar{s}$  state as f$_1$(1427) while
for the $1^{+-}$ meson we use  the ansatz that $M^2_{s\bar{s}} -
M^2_{q\bar{q}}  \approx 0.5 $ GeV$^2$ to set the $s\bar{s}$ state at
1.42 GeV using the observed b$_1$ mass. Similarly we estimate the 
$s\bar{s}$ state for $2^{-+}$ at 1.81 GeV using the observed  $\pi_2$
mass.

For the scalar mesons, the experimental situation is complex - possibly 
owing to mixing with glueballs and meson-meson thresholds. As an estimate
of the $s\bar{s}$ mass we again use the prescription of adding 0.5~GeV$^2$
to the a$_0$ mass squared yielding an estimate of 1.21 GeV.

We have presented an estimate of the  experimental spectrum for  $s
\bar{s}$ quarks. We now need to justify that this is appropriate  for
our lattice study.  We measure the vector to pseudoscalar mass ratio as
$M_V/M_P=1.31(1)$  and use the value of $J={1 \over 2}dM_V^2/dM_P^2$ to
relate this ratio to the experimental  K$^{*}$ to K mass ratio. The
average quark mass in  a pseudoscalar meson is  then assumed to be
proportional to $M_P^2$. Values of $J$ from quenched lattices~\cite{J}
are  near 0.37, while  the observed meson  spectrum yields values of
0.44 - 0.48. This uncertainty leads to an  estimate of our quark mass as
1.3 to 1.5 times the strange quark mass. We present our results as
ratios to  the vector meson mass and the systematic error from treating
our quark  mass as that of the strange quark is quite small for the
$L$-excited and hybrid  mesons compared to other sources  of error.

Since the experimental states are for full QCD, they include decay
effects  which may also shift the mass somewhat from the quenched value.
Some evidence for departures  of the quenched meson spectrum from that
of experiment has already  been claimed~\cite{J} from studying the quark
mass dependence. Here we are able to extract masses for a wide range of
mesons. In  this initial study, we use a fixed lattice spacing so we are
unable to  extract continuum limits.  Nevertheless, the splittings
between similar  states (such as the $P$-wave mesons) would be expected
to be reliably  given at our lattice spacing. We work with quarks of
fixed mass (close to the $s$ quark mass) and present our lattice results
as  ratios of masses to the vector meson mass ($\phi$). As shown in
Table~1, our  overall spectrum is close to that of experiment (as
interpreted above - see Table~1). We now  discuss various topics in more
detail.

We are able, for the first time, to  measure a non-zero mass difference 
among the $P$-wave states: the $J$=0 member of the spin-triplet   is
found to be lighter. The sign of this splitting is the same as that
found  experimentally for $c\bar{c}$ mesons where the $J$=0 state  is
significantly lighter than the other $P$-wave states.  The magnitude of
the splitting of the $0^{++}$ meson from the centre of gravity of the
$P$-wave  mesons  is somewhat smaller than our estimate from  experiment
for $s\bar{s}$ quarks. As discussed above, the experimental scalar meson
spectrum is  difficult to interpret and our quenched results can act as
a guide to  phenomenological quark models.

For the $D$-wave states, we are able to estimate the masses but with quite
large statistical and systematic errors.  The mass values we  obtain
appear to be somewhat higher than those expected experimentally. 
Bearing in mind the need to  extrapolate to $a=0$, the corrections could
be large since, at our lattice spacing, $ma > 1 $  for these states. 
Another possible problem is that, since $ma$ is  large, the signal
disappears  quickly with increasing $t$ and so excited state
contributions may not be completely removed. Our method does, however,
enable us to  get meaningful signals and a route to more precision is
available  using higher statistics and a smaller lattice spacing.

Previous to this work, the only lattice determination of the  hybrid
meson spectrum  has been  that obtained~\cite{hybrid}  in the static
limit of the quenched approximation.  The static quark analysis showed
that the lowest  lying gluonic excitation of the potential $V(R)$ has
the symmetry  $\, \sqcap - \sqcup$. Assuming the adiabatic
approximation, the bound states in this  potential will have a lowest
excitation energy with one unit of  angular momentum about the quark -
antiquark separation axis. These  excitations then correspond to 
$L^{PC}=1^{+-}$ and $1^{-+}$ spatial  wavefunctions and are degenerate
in energy,  as described in the Appendix. Thus  all eight hybrid mesons
would be degenerate in this approach. 

Within the static quark framework, this degeneracy will be broken  by
spin-orbit effects which shift the spin-exotic levels and cause mixing 
of the non-exotic levels with ordinary $q\bar{q}$ states. One clue to 
the sign of this effect for static quarks comes from considering  the
limit of zero separation. In this case of $r=0$,  the symmetry
classification  of the gluonic states is different from that for  $r \ne
0$, and the energies  of these states  have also been measured on a
lattice~\cite{gluelump} with the result that the $1^{+-}$ gluonic mode
is several hundred MeV lighter than the $1^{-+}$ mode.  This, in turn,
suggests that hybrid states with  the $1^{+-}$ spatial wavefunction are
likely to be lighter: so that  $J^{PC}=1^{-+}$  would be the lightest
spin exotic hybrid.

We are able to explore this  splitting of the degeneracy by a completely
different  method - explicitly studying each state on a lattice using
propagating quarks. We construct operators with  $1^{+-}$ and $1^{-+}$
spatial wavefunctions, couple them to quark bilinears  and measure the
masses of the mesonic states.  Our present results for the masses of the
spin-exotic states are not sufficiently accurate to indicate the level 
ordering unambiguously.  One theoretical clue is that as $r \to 0$, our
$1^{-+}$ spatial wavefunction decouples faster than the $1^{+-}$ spatial
wavefunction.  This analysis at $r=0$ implies in turn that hybrid states
with a $1^{+-}$ gluonic mode are likely to be at lower energy, resulting
in a splitting of the hybrid states with  $J^{PC}=1^{-+}$ being the
lightest exotic meson, as argued above for the case  of static quarks. 
Our results are consistent with this scenario.

We are also able to measure matrix elements between our hybrid operators
and the ground state mesons for the non-exotic $J^{PC}$ states. Our
conclusion  is that the mixing is surprisingly strong. This would imply
that hybrid  levels with pseudoscalar and vector quantum numbers would
be unlikely  to exist unmixed in nature.

There have been several experimental claims for  hybrid   mesons - for a
review see ref~\cite{close}.  Our results suggest that  non spin-exotic
candidates may need re-appraisal since big mixing effects  are possible.
For the exotic  mesons, the favoured candidate to lie lowest will  have
$J^{PC}=1^{-+}$ and several experimental hints of such states have  been
reported.  

If our discussion of the quark mass dependence  of states in the
quenched approximation applies to hybrid mesons, then  we may use  the
ansatz that $M^2_{s\bar{s}} - M^2_{q\bar{q}}  \approx 0.5 $ GeV$^2$
where $q$ means $u$ or $d$. This suggests that hybrid mesons composed of
$u$ and $d$  mesons will be approximately  150~MeV lighter than the $s
\bar{s}$ mesons for masses around 1.8 GeV.  For strange mesons, the lack
of C invariance will mask the experimental identification of spin
exotic states.

We have presented methods which enable the meson spectrum to be  fully
explored for all $J^{PC}$ values using non-perturbative lattice methods.
The results we have obtained show that accurate determinations of masses
are  possible by these methods. We intend to extend this approach by 
using more statistics, several quark mass values and a range of lattice 
spacings so as to get precise continuum masses in the quenched
approximation.

\section{Acknowledgements}

We acknowledge support from EPSRC grant GR/K/41663 to the UKQCD 
Collaboration.  We thank  David Richards for his help  in setting up the
inversion codes for the shifted local propagators.

\appendix

\section{Appendix: Construction of mesonic operators.}

To study spectroscopy in lattice gauge theory, one needs operators at 
fixed time which create and destroy a mesonic state. By considering the 
correlation of such operators at time separation $t$ in the vacuum,  the
masses of the transfer matrix eigenstates can be extracted. We consider
the construction of bosonic operators which contain a quark and an
antiquark. The case we shall explore is where the quark is at $x_1$ and
antiquark  at $x_2$ and they are joined by a product of links along a
path P. This is a gauge invariant construction. The general form of
the operator is then  
 $$
  \bar{\psi}(x_2) \ \Pi_{\rm P} U \ \Gamma \psi(x_1)
$$ 
 where $\Gamma$ is one of the 16 independent $\gamma$-matrix
combinations.  We shall be concerned with constructing  operators 
with a given set of quantum numbers by taking linear combinations 
with different paths and different expressions for $\Gamma$.

For orientation, we first consider the case where there is no path P
and the local operator has $x_1=x_2$.  Then in the continuum such a
bilinear can have $J^{PC}$ values as listed.  The historic name of the
isovector non-strange meson with those  quantum numbers is also added
for convenience.

\begin{tabular}{ccc}
 $\Gamma$  & $ J^{PC}  $&   name  \\
  1        &$   0^{++} $&$    \delta$ \\
$  \gamma_4 $&$   0^{+-} $&    {\it exotic} \\
$  \gamma_i $&$   1^{--} $&$    \rho$ \\
$  \gamma_i \gamma_4 $&$  1^{--}$&$ \rho$ \\
$  \gamma_5 $&$   0^{-+}   $&$ \pi$ \\
$  \gamma_5 \gamma_4 $&$0^{-+} $&$\pi $ \\
$  \gamma_i \gamma_j $&$1^{+-} $&B \\
$  \gamma_5 \gamma_i $&$1^{++} $&A$_1$ \\
\end{tabular}

\noindent Note that the non-relativistic limit of these 16 bilinears,
which corresponds  to projecting the quark spinor with $1+\gamma_4$,
leaves 4 combinations which have  the $\pi$ and $\rho$ quantum  numbers
only - as in the naive quark model. Note also that when the quark and
antiquark  have different masses, the meson will not be an eigenstate of
charge  conjugation C. Thus, for example, the $1^{+-}$ and $1^{++}$
axial mesons will mix.

The general case will involve discussion of the properties of the path P
under rotations, reflections and charge conjugation. Since the spinor
structure is in a  fixed basis, it is sufficient to consider separately
the transformation properties of the path P  and then combine the result
with that given above for the quark bilinears. With this in mind, we now 
consider the classification of the path P under the symmetries of the 
lattice: rotation, reflection, translation and charge conjugation.

\subsection{Discrete group theory}

The group of rotations and inversions for a three dimensional 
spatial lattice is the cubic group $O_h$.  This is thus the 
appropriate classification group for bosonic transfer matrix eigenstates 
with momentum zero.  For non-zero momentum the space group 
needs to be used instead - see ref~\cite{space}.

Here we gather together some of the appropriate properties of  this
group and its representations.  The representations are  labelled by
parity P and charge conjugation C in the same way as bosonic
representations of the SU(2) rotation group appropriate to  the
continuum formulation. Thus the essential difference between the lattice
eigenstates and continuum eigenstates lies in the `spin'. 

Whereas a state of spin $J$ has 2$J$+1 spin components which are degenerate
in mass and form a 2$J$+1 dimensional representation in the continuum, 
for $O_h$ there exist only 1, 2 and 3-dimensional representations.
The 1-dimensional representations are labelled $A_1$ and $A_2$, there 
is a 2-dimensional $E$ representation and the 3-dimensional 
representations are $T_1$ and $T_2$. 

The relationship of these representations to those of SU(2) can be 
derived by restricting the SU(2) representations to the rotations 
allowed by cubic symmetry and classifying them under $O_h$. This 
process (called subducing) yields the results (tabulated to $J$=4):

\begin{tabular}{cc}
$J$=0 &$ A_1$ \\
$J$=1 &$ T_1$ \\
$J$=2 &$ E \ T_2$ \\
$J$=3 &$ A_2 \  T_1\  T_2$ \\
$J$=4 &$ A_1 \  E \ T_1 \  T_2$ \\
\end{tabular}

Thus, for example,  as the lattice spacing is decreased, a $J$=2 state
will be  recognised by degenerate masses in $E$ and $T_2$
representations. This  is indeed observed in glueball
studies~\cite{oh,liv}. Conversely, a state observed in the $T_2$
representation  on a lattice could have $J$=2 or 3 and this choice can
only be resolved  by observing the degenerate partners (e.g. $E$ for $J$=2
or $A_2$ and $T_1$  for $J$=3). In this way, with sufficiently accurate 
lattice spectra in all $O_h$ representations, the rotationally
invariant continuum levels  can be constructed for any $J$ value in
principle.

Optimal signal to noise comes from constructing  operators  on a lattice
which  create states of given $O_h$ representation. This has been well
documented  in glueball studies where the appropriate observables are
closed Wilson loops - see ref~\cite{cmAP}.  In the present study, the
appropriate lattice constructs are the  paths P from $x_1$ to $x_2$
along the links of the lattice. These paths  must then be classified
according to the representations of the cubic group $O_h$.  The most
powerful method is to use the projection table~\cite{cmAP}  which
directly gives the relevant combination of paths with different cubic
rotations for each representation.

As well as the spatial path P, the mesonic operator will have  a spin
component coming from the quark spinors. On a lattice, where the spatial
symmetry is $O_h$,  the only change in  the classification of these
bilinears is that the singlets ($J=0$) become $A_1$   while the triplets
($J=1$) become $T_1$. This change has one implication which has not been
discussed before.  The `$\rho$' bilinear on a lattice is in the $T_1$
representation and so  allows $J$=1 and 3. Thus the excited state of the
$\rho$ meson seen in lattice studies could  be the g(3$^{--}$) rather
than a heavier $\rho(1^{--})$ as usually assumed.

We need to combine the $O_h$ representation of this  quark bilinear with
that of the  spatial path P (from $x_1$ to $x_2$). The combined operator
thus will lie in a representation  given by the Clebsch-Gordan
decomposition of the product of representations. When the quark bilinear
is in  the trivial $A_1$ representation, this is straightforward since
the final representation  is just that of the spatial path. For the 
$T_1$ representation of quark bilinears,  we need the decomposition:

\begin{eqnarray*}
 T_1 \otimes A_1 &=& T_1   \nonumber \\
 T_1 \otimes A_2 &=& T_2   \nonumber \\
 T_1 \otimes E\  &=& T_1 \oplus T_2   \nonumber \\
 T_1 \otimes T_1  & =& A_1 \oplus T_1 \oplus T_2 \oplus E   \nonumber \\
 T_1 \otimes T_2  & =& A_2 \oplus T_1 \oplus T_2 \oplus E   \nonumber \\
\end{eqnarray*}

As well as the Clebsch-Gordan decomposition, we will need the
Clebsch-Gordan coefficients to  construct the explicit operators
required. These will be given later for the cases of most interest.

Since the general path with no symmetry will contribute to all $O_h$ 
representations in principle, one can create operators with all 
quantum numbers. Since it is practical to start with simpler 
cases, we now discuss the states that can be constructed from 
more symmetric paths: straight, L-shaped and U-shaped.

\subsection{Orbitally excited mesons.}

A full study of mesons with orbital angular momentum needs  non-local
operators. The  simplest case to consider is that of a straight line
from $x_1$ to $x_2$ joining the quark and antiquark. If this  straight
line is along  the lattice axis then the result is that there are $A_1$,
$T_1$ and $E$  representations. In a notation where the directed path is
labelled by the axis  of its orientation these representations 
are  given explicitly by:

\begin{tabular}{cc}
$A_1^{++} $&$ s_1+s_2+s_3$\\
$T_1^{--} $&$  p_1,\ p_2,\ p_3 $\\
$E^{++} $&$ s_1-s_2,\ 2s_3-s_1-s_2 $\\
\end{tabular}

\noindent where $p_j=(j-\bar{j})$ and $ s_j=(j+\bar{j})$ and where
$\bar{1}$ is the path directed along the negative $x$-axis, for
example.  The conventional normalisation factors have been omitted
to simplify the notation.

These three representations correspond partly to the $S$-, $P$- and
$D$-wave orbital excitations of the naive quark model. The path
combinations  given play the role of the spherical harmonics in SU(2). A
discussion of the construction of operators in the $T_1^{--}$
representation has also been given by ref.~\cite{wingate} in  their
study of $P$-wave states.

Consider now the states that can be constructed by combining these  path
representations with the quark bilinears. For the `large components'
which survive in the non-relativistic limit,  the bilinears are in
representations $A_1^{-+}$ and $T_1^{--}$. Combining these with the
above straight-path representations yields:

\begin{tabular}{ccccc}
$\bar{q} \Gamma q $&  Path  &  State  &     operator& $J^{PC}$\\
& & & & \\
$  A_1^{-+}  $&$      A_1^{++}$&$ A_1^{-+} $&
    $  \gamma_5 (s_1+s_2+s_3)$&$ 0^{-+},\ 4^{-+}$\\
$  T_1^{--}  $&$     A_1^{++} $&$T_1^{--}  $&
   $ \gamma_j (s_1+s_2+s_3)$&$ 1^{--},\ 3^{--}$\\
 & & && \\
$  A_1^{-+}  $&$     T_1^{--} $&$T_1^{+-} $&
   $ \gamma_5 p_j $&$ 1^{+-},\ 3^{+-} $\\
$  T_1^{--}   $&$     T_1^{--}$&$ A_1^{++}$& 
   $  \gamma_1 p_1+\gamma_2 p_2+\gamma_3 p_3 $&$ 0^{++},\ 4^{++}$\\
$  T_1^{--}   $&$    T_1^{--} $&$T_1^{++} $&
   $  \gamma_k p_i-\gamma_i p_k$&$ 1^{++},\ 3^{++}  $\\
$  T_1^{--}   $&$    T_1^{--} $&$T_2^{++} $&
   $  \gamma_k p_i+\gamma_i p_k $&$ 2^{++},\ 3^{++}$ \\
$  T_1^{--}   $&$    T_1^{--}$&$  E^{++}  $&
  $  \gamma_1 p_1-\gamma_2 p_2,
  \ 2\gamma_3 p_3-\gamma_1 p_1-\gamma_2 p_2 $&$ 2^{++},\ 4^{++} $ \\
 & & & & \\
$  A_1^{-+} $&$      E^{++}$&$   E^{-+}   $&
     $\gamma_5 (s_1-s_2),\ \gamma_5 (2s_3-s_1-s_2) $&$ 2^{-+},\ 4^{-+} $ \\
$  T_1^{--} $&$      E^{++}$&$ T_1^{--}   $&
   $\gamma_j (2s_j-s_i-s_k) $&$ 1^{--},\ 3^{--} $\\
$  T_1^{--} $&$      E^{++}$&$ T_2^{--}   $&
   $\gamma_j (s_i-s_k) $&$ 2^{--},\ 3^{--} $\\
\end{tabular}

\noindent where $i,\ j,\ k$ are cyclic and there is no implied 
summation.   The two lowest $J$-values corresponding to the $O_h$
representation are also given for convenience. The above operators have
the same quantum numbers  when multiplied by $\gamma_4$, indeed
projection with $(1+\gamma_4)/2$  is useful to reduce computation. Note
that while operators with different quantum numbers are  orthogonal to
each other, the two $T_1^{--}$ operators will mix through spin-orbit
interactions.

Note that with an isotropic spatial distribution,  the  quark bilinear
can create, through  the `small' spinor components, the $P$-wave states
$A_1^{++}, \  T_1^{+-}$ and $T_1^{++}$  (particles $\delta$,  B and
A$_1$). These extra operators allow a fuller variational  basis to be
used in the study of these particular states.

The lowest $J^{PC}$ states created are just those of the continuum
$L$-excited naive quark model for $S$- and $P$-waves with an incomplete
$D$-wave in the sense that the $D$-wave mesons with $J^{PC}=2^{--}$ and
$3^{--}$ will both  be created by the $T_2^{--}$ operator above. In
order to separate these  states, one needs to use operators in the
$E^{--}$ and $A_2^{--}$  representations. One way in which these are
accessible  is by using a `straight diagonal path' from quark to
antiquark.  Such a path is the sum of the two L-shaped paths going to
the far corner of a square via the sides. It acts, in some way, as a
lattice opportunity to  study operators at 45$^0$ to the lattice axes
and so helps to explore  higher spin. The 12 such paths contribute to 
the $A_1^{++},\ E^{++},\ T_2^{++},\ T_2^{--}, \ T_1^{--} $
representations. Using a notation $[i\bar{j}]$ to describe such a path
to  the $(i,-j)$ diagonal corner from the origin etc., then appropriate
combinations for the $T_2^{++}$ spatial operator are 
 $$ 
t_k= [ij] -[i\bar{j]}-[j\bar{i]}+[\bar{i}\bar{j}]
 $$
 \noindent with $i,\ j,\ k$ cyclic. Combining this with the $T_1^{--}$
quark bilinear gives the required operators:

\begin{tabular}{ccccc}
$\bar{q} \Gamma q $&  Path  &  State  &     operator& $J^{PC}$\\
 & & & \\
$  T_1^{--} $&$      T_2^{++}$&$ A_2^{--}   $&
  $ \gamma_1 t_1+\gamma_2 t_2+\gamma_3 t_3 $&$ 3^{--},\ 5^{--} $\\
$  T_1^{--} $&$      T_2^{++}$&$ E^{--}   $& $\gamma_1 t_1-\gamma_2 t_2,\ 
     2 \gamma_3 t_3 -\gamma_1 t_1-\gamma_2 t_2$&$ 2^{--},\ 4^{--} $\\
$  T_1^{--} $&$      T_2^{++}$&$ T_1^{--}   $&
   $ \gamma_i t_k+\gamma_k t_i $&$ 1^{--},\ 3^{--} $\\
$  T_1^{--} $&$      T_2^{++}$&$ T_2^{--}   $& 
    $ \gamma_i t_k-\gamma_k t_i $&$ 2^{--},\ 3^{--} $\\
\end{tabular}

This illustrates that the $2^{--}$ and $3^{--}$ states can be separated 
and thus explored  from these operators.  This discussion of straight
paths completes the construction of operators which allow a full study
of the $D$-wave mesons. Our results for  the lattice operators needed to
construct a $D$-wave $\rho$ meson are similar to those presented in
ref~\cite{gupta}.

It is possible to study some aspects of $F$-wave mesons on the lattice
fairly simply. For example, consider the `straight cubic diagonal'
paths which go from the origin to the corners of a cube  centered at
the origin. Paths with this symmetry can be defined  on a lattice as a
sum over the six equivalent shortest routes to a corner along the
links in each case.  From such path combinations, the $A_2^{--}$
representation  can be constructed as a sum of these paths to each of
the 8 corners of the cube with a sign that alternates between adjacent
corners. This represents a spatial $L$=3 excitation. When  combined with
the $A_1^{-+}$ quark bilinear, this yields a $A_2^{+-}$ state which
enables study of  the $3^{+-}$ F-wave spin-singlet  meson.

\subsection{Hybrid and exotic $J^{PC}$ states}

Hybrid mesons are defined as having the gluonic field excited in  a
non-trivial way. In the case of static quarks, this is a very  clear
prescription and a thorough study has been made of the  excited gluonic
energy levels~\cite{hybrid}. Solving for the meson spectrum in this 
excited static potential in the adiabatic approximation then gives 8
degenerate lowest  lying  hybrid states.  One of the  consequences of
these excited gluonic modes is that hybrid mesons  can have $J^{PC}$
values not allowed to mesons which have their  gluonic  degrees of
freedom in the ground state. We are particularly interested in
constructing mesonic operators with $J^{PC}$-values which are  not
present in the naive quark model  since these have a clear signature
experimentally. These have {\it exotic} quantum numbers  which are
$0^{--}, 0^{+-}, 1^{-+}, 2^{+-}, $ etc.  Because of the many-to-one map
from $O_h$ representation to $J^{PC}$, it is necessary in principle to
study all $O_h$ representations as discussed above. Nevertheless,
promising representations to study are $A_1^{--},\ A_1^{+-},\ T_1^{-+}$
and $E^{+-}$ since they only have  a non-exotic contribution at
$J$-values 3 or more  higher  than the exotic one - unlike the case of
the $T_2^{+-}$ representation where the exotic $J$=2 state may be
contaminated by  contributions from a non-exotic $J$=3 meson.

For straight paths, none of the states in the table above are of the 
required exotic representation. This is to be expected since the
straight  path corresponds to the naive $L$-excited quark model and so
should not  generate exotic $J^{PC}$ states. The exception is that there
are  relativistic couplings  to quark bilinears which do allow exotic 
representations.  If one has in mind the picture of a hybrid state  with
exotic quantum numbers as arising from gluonic excitations, then  such
operators with no gluonic excitation are not particularly promising in
the sense that the  coupling may be very  small and hence swamped by
noise in practice.

The more direct way to create exotic states is to use a path from $x_1$ 
to $x_2$ which  is not straight.  The evidence from a lattice study of 
hybrid mesons  formed from static quarks~\cite{hybrid} is that  the
lowest energy excitation comes from U-shaped paths.  In our case,  a
U-shaped path will have 24 different orientations and will contribute to
the following  representations: $A_1^{++},\ A_2^{++},\ E^{++}$ (twice),
$\ T_1^{--},\ T_1^{+-},\ T_1^{-+},\ T_2^{--},\ T_2^{+-},\ T_2^{-+}$.
Combining these spatial representations with the quark bilinears 
$A_1^{-+}$ and $T_1^{--}$ then gives access to most of the exotic states
discussed above as well as many copies of the non-exotic ones.

 The static quark results show~\cite{hybrid} that the lowest energy
hybrid states come from an  operator which is the difference of 
U-shaped paths from quark to antiquark of the form $\, \sqcap - \sqcup$.
We find that in this study also, exotic states are  only created from
this  difference of paths and so we specialise to this case. The
$k$-component of the representation of the spatial paths  is  given by
the following (where $ij$ refers to the path combination  $\, \sqcap -
\sqcup$ from $x_1$ to $x_2$ along the $i$-axis with  transverse extent
in the $j$-direction);

\begin{tabular}{cc}
$T_1^{+-} $&$ u_k=ij-ji-\bar{i}j+\bar{j}i$\\
$T_1^{-+} $&$ v_k=ik+\bar{i}k+jk+\bar{j}k$\\
$T_2^{+-} $&$ U_k=ij+ji-\bar{i}j-\bar{j}i$\\
$T_2^{-+} $&$ V_k=ik+\bar{i}k-jk-\bar{j}k$\\
\end{tabular}

\noindent with $i,\ j,\ k$ cyclic.  The $T_1$ spatial path combinations 
are illustrated here:

\begin{figure}[h]
\setlength{\unitlength}{0.6mm}
\begin{picture}(100,40)
\put(20,10){\line(0,0){20}}
\put(10,20){\line(1,0){20}}
\put(20,30){\vector(-1,0){20}}
\put(10,20){\vector(0,-2){20}}
\put(20,10){\vector(1,0){20}}
\put(30,20){\vector(0,0){20}}
\put(0,20){\line(0,0){10}}
\put(40,10){\line(0,0){10}}
\put(20,40){\line(1,0){10}}
\put(10,0){\line(1,0){10}}
\large
\put(50,20){$-$}
\put(45,0){$T_1^{+-}$}
\normalsize
\put(80,10){\line(0,0){20}}
\put(70,20){\line(1,0){20}}
\put(80,10){\vector(-1,0){20}}
\put(90,20){\vector(0,-2){20}}
\put(80,30){\vector(1,0){20}}
\put(70,20){\vector(0,0){20}}
\put(60,10){\line(0,0){10}}
\put(100,20){\line(0,0){10}}
\put(70,40){\line(1,0){10}}
\put(80,0){\line(1,0){10}}

\put(120,15){\line(0,0){10}}
\put(140,15){\line(0,0){10}}
\put(140,25){\vector(-1,0){20}}
\put(150,15){\line(0,0){10}}
\put(170,15){\line(0,0){10}}
\put(150,25){\vector(1,0){20}}
\put(124,4.5){\line(0,0){10}}
\put(141,13){\line(0,0){10}}
\put(141,23){\vector(-2,-1){17}}
\put(149,27){\line(0,-2){10}}
\put(166,35.5){\line(0,-2){10}}
\put(149,27){\vector(2,1){17}}
\large
\put(180,20){$-$}
\put(175,0){$T_1^{-+}$}
\normalsize

\put(190,15){\line(0,0){10}}
\put(210,15){\line(0,0){10}}
\put(210,15){\vector(-1,0){20}}
\put(220,15){\line(0,0){10}}
\put(240,15){\line(0,0){10}}
\put(220,15){\vector(1,0){20}}
\put(194,4.5){\line(0,0){10}}
\put(211,13){\line(0,0){10}}
\put(211,13){\vector(-2,-1){17}}
\put(219,27){\line(0,-2){10}}
\put(236,35.5){\line(0,-2){10}}
\put(219,17){\vector(2,1){17}}

\end{picture}
\end{figure}

The $T_1$ spatial representations  correspond to $L$=1 and so should lie
lower in energy than the  $T_2$ cases. Thus we expect the lowest lying
hybrid mesons to be  obtained by combining the $T_1$ spatial behaviour
with the quark  bilinear. This gives the  operators shown in Table~3.
The mesonic quantum numbers of these hybrid states are exactly the 
same as those produced by the lowest gluonic excitation for static 
quarks~\cite{hybrid}. This includes most of the exotic possibilities.

\begin{table}[h]
\begin{tabular}{ccccc}
$\bar{q} \Gamma q $&  Path  &  State  &     operator & $J^{PC}$\\
& & & & \\
$  A_1^{-+} $&$  T_1^{+-}$&$ T_1^{--}   $&
   $\gamma_5 u_j $&$ 1^{--},\ 3^{--} $\\
$  T_1^{--} $&$  T_1^{+-}$&$ A_1^{-+}   $&
   $\gamma_1 u_1+\gamma_2 u_2+\gamma_3 u_3 $&$ 0^{-+},\ 4^{-+} $\\
$  T_1^{--} $&$  T_1^{+-}$&$ E^{-+}   $&
   $\gamma_1 u_1-\gamma_2 u_2,\  
   2\gamma_3 u_3-\gamma_1 u_1-\gamma_2 u_2 $&$ 2^{-+},\ 4^{-+} $\\
$  T_1^{--} $&$  T_1^{+-}$&$ T_1^{-+}   $&
   $\gamma_k u_i-\gamma_i u_k $&$ 1^{-+},\ 3^{-+} $\\
$  T_1^{--} $&$  T_1^{+-}$&$ T_2^{-+}   $&
   $\gamma_k u_i+\gamma_i u_k $&$ 2^{-+},\ 3^{-+} $\\
&&&&\\
$  A_1^{-+} $&$  T_1^{-+}$&$ T_1^{++}   $&
   $\gamma_5 v_j $&$ 1^{++},\ 3^{++} $\\
$  T_1^{--} $&$  T_1^{-+}$&$ A_1^{+-}   $& 
 $\gamma_1 v_1+\gamma_2 v_2,+\gamma_3 v_3  $&
   $ 0^{+-},\ 4^{+-} $\\
$  T_1^{--} $&$  T_1^{-+}$&$ E^{+-}   $& 
 $\gamma_1 v_1-\gamma_2 v_2, \ 2\gamma_3 v_3-\gamma_1 v_1-\gamma_2 v_2   $&
   $ 2^{+-},\ 4^{+-} $\\
$  T_1^{--} $&$  T_1^{-+}$&$ T_1^{+-}   $&
   $\gamma_k v_i-\gamma_i v_k $&$ 1^{+-},\ 3^{+-} $\\
$  T_1^{--} $&$  T_1^{-+}$&$ T_2^{+-}   $&
   $\gamma_k v_i+\gamma_i v_k $&$ 2^{+-},\ 3^{+-} $\\
\end{tabular}
 \caption{Hybrid meson operators, where $i,\ j,\ k$ are cyclic and factors
of $\gamma_4$ can be included too.}
 \end{table}

As a further support for this  identification of  the lowest lying
hybrids, we can compare with hybrid operators  constructed using
L-shaped paths instead of U-shaped.  Considering this time  the
difference of the two L-shaped paths to the opposite corner of a square
via the edges,   this combination contributes  to $E^{+-},\ A_2^{+-},\
T_1^{+-},\ T_1^{-+},\ T_2^{-+}$ spatial representations. This set of
hybrid spatial excitations is very similar to that from  U-shaped
paths. In particular the  excitations of lowest $L$ ($T_1$ with $L$=1)  are
the same. For completeness we give the construction of these  hybrid
spatial paths, using a notation that the L-shaped path first in the
$i$-direction  and then in the $j$-direction is $(ij)$ with $\bar{i}$
representing the  negative direction etc., with $i,\ j,\ k$ cyclic:  
 \begin{eqnarray*}
 T_1^{+-} \ \ \ \ & (ij)-(ji)+(j\bar{i})-(\bar{i}j)+
  (\bar{i}\bar{j})-(\bar{j}\bar{i}) +(\bar{j}i)-(i\bar{j})
\nonumber \\
 T_1^{-+}  \ \ \ \ &(ki)+(k\bar{i})+(kj)+(k\bar{j}) 
     - (\bar{k}i)-(\bar{k}\bar{i})-(\bar{k}j)-(\bar{k}\bar{j}) 
\nonumber \\
    &  +(i\bar{k})+(\bar{i}\bar{k})+(j\bar{k})+(\bar{j}\bar{k})
      -(ik)-(\bar{i}k)-(jk)-(\bar{j}k) 
      \nonumber \\
 \end{eqnarray*}

One exotic state which  is of particular interest is the $A_1^{+-}$
state  since it can also be created from an isotropic spatial
construction by a  quark bilinear using $\Gamma=\gamma_4$. Thus one can
explore  its creation  and annihilation using either the isotropic
operators (local or $S$-wave)  or the non-local operators (U or
L-shaped). Again the isotropic spatial construction does not have an 
excited gluonic field and so may not give good signal to noise for an
exotic  hybrid meson.

Only one  exotic representation $(A_1^{--})$ cannot be obtained using
the paths discussed above. To create this state needs paths  with even
less symmetry than the planar U and L-shaped cases we have discussed in
detail.  Because such a complex path shape is necessary, one expects that
the state  will be relatively high lying in mass.

For completeness, we point out that gluonic excitations of light quarks
can be studied in principle in the limit $x_1 \to x_2$ where only local
quark propagators will be needed. The above U- and L-shaped paths are
then not appropriate and the simplest choice is a square path with one
corner at the quark and  antiquark source. The construction of
combinations of such paths in appropriate  representations of the cubic
group is discussed in  ref~\cite{gluelump} where they were used with a
static adjoint colour source. Here the  same construction is needed and
the quark and antiquark at $x_1=x_2$ are joined by the square paths. The
lowest gluonic excitations are  expected to come from the $T_1^{+-}$ and
$T_1^{-+}$ spatial operators  again - just as for $x_1 \ne x_2$. These 
spatial operators are then to be combined with quark bilinears just as 
above.

\subsection{Mesonic correlations.}

The mesonic operators we have discussed are of the form
 $$
  \bar{\psi}_2(x_2) \ P(x_2,x_1)  \ \Gamma \ \psi_1(x_1)
$$ 
 where $\Gamma$ is one of the 16 independent $\gamma$-matrix
combinations and $P$ is a product of links.  The subscripts  on the
spinors refer to the possibility of different masses  for quark and
antiquark. A sum over $x_1$ and $x_2$ with appropriate path combinations
is needed to  project out the quantum numbers of interest as discussed
above. We consider the operator defined in a time slice  at time $t$.

The mesonic state can be explored on a lattice by measuring  the
correlation of this operator at $t=0$ with a similar operator  at  time
$t$.  Allowing for different operators at  source (A, $x$ with $t=0$)
and sink (B, $y$ with time $t$) then gives the correlation to be 
measured
$$
 C_{BA}(t)= 
 < \bar{\psi}_1(y_1) \gamma_4 (P^B \Gamma^B)^{\dag} \gamma_4 \psi_2(y_2)
\bar{\psi}_2(x_2) P^A \Gamma^A \psi_1(x_1)>
$$
 In the quenched approximation, for a Wilson-like fermionic action, the
quark propagators $S$ can be  explicitly introduced to re-express this
as
 $$
 C_{BA}(t) =-< \gamma_5 S_1^{\dag}(y_1,x_1) \gamma_5 \gamma_4
 (P^B \Gamma^B)^{\dag} \gamma_4 S_2(y_2,x_2) P^A \Gamma^A >
$$
  For A=B, this correlation is positive definite.

This expression for the correlation has the colour structure of two
propagators connected  by paths P$^A$ and P$^B$.

The symmetries of the lattice can be used to increase the precision 
of a measurement by combining it with the correlation measured on 
lattices reflected in the time or space directions using the 
identity for the propagators that
$$
   S(x,y;U) =\gamma_i \gamma_5 S(x^I,y^I;U^I) \gamma_5 \gamma_i
$$
 where the superscript $I$ refers to the lattice coordinates and links
after  inversion in the $i$ direction (where $i=1,\ 2,\ 3$ or 4). 
Charge conjugation provides another relation since
$$
   S(x,y;U) =\gamma_4 \gamma_2 \gamma_5 S^*(x,y;U^*) 
  \gamma_5 \gamma_2 \gamma_4
$$

Note  that the identity  for inversion with  $i=4$ can be used to
combine measurements of the correlation at $t= \pm |t|$  using
$C_{BA}(-t)=\pm C_{BA}(t)$ with the appropriate sign coming from
commuting  $\gamma_4 \gamma_5$ past  $\Gamma^A$ and $\Gamma^B$.

In practice we construct the sink operator (B) to have the required
quantum  numbers but employ a source operator (A) which has less
symmetry but a non-zero  overlap with the sink operator. In this
situation, it is useful to  use the symmetries of the lattice under
space reflections to eliminate  terms in the correlation which
have  zero expectation value. This  reduces the statistical noise of
the measurement.

Note that for the U-shaped paths, even though we  use a fixed source 
direction, we  measure correlations of source and sink  operators for 64
different combinations of orientations and gamma matrices. After
averaging equivalent cases,  these correlations allow  a reconstruction
of the mesonic quantum  numbers of the hybrid states tabulated above.

\end{document}